\documentclass[%
 reprint,
superscriptaddress,
 amsmath,amssymb,
 aps,pre,hyphens, andfloatfix
]{revtex4-2}

\usepackage{hyperref}
\hypersetup{hypertex=true,
colorlinks=true,
linkcolor=blue,
anchorcolor=blue,
citecolor=blue}
\usepackage{enumerate}
\usepackage[usenames,dvipsnames,svgnames,table]{xcolor}
\usepackage{IEEEtrantools}
\usepackage{graphicx}
\usepackage{array}
\usepackage{multirow}
\usepackage{verbatim}
\usepackage{float}
\usepackage{amsmath}
\usepackage{subfigure}
\usepackage{blindtext}

\usepackage{xcolor}
\usepackage{marginnote}

\begin{document}

\preprint{APS/123-QED}

\title{Supervised and unsupervised learning of (1+1)-dimensional even-offspring branching annihilating random walks}

\author{Yanyang Wang}
\affiliation{Key Laboratory of Quark and Lepton Physics (MOE) and Institute of Particle Physics, Central China Normal University, Wuhan 430079, China}

\author{Wei Li}
\email[]{liw@mail.ccnu.edu.cn}
\affiliation{Key Laboratory of Quark and Lepton Physics (MOE) and Institute of Particle Physics, Central China Normal University, Wuhan 430079, China}

\author{Feiyi Liu}
\affiliation{Key Laboratory of Quark and Lepton Physics (MOE) and Institute of Particle Physics, Central China Normal University, Wuhan 430079, China}
\affiliation{Institute for Physics, E{\"o}tv{\"o}s Lor\'and University\\1/A P\'azm\'any P. S\'et\'any, H-1117, Budapest, Hungary}

\author{Jianmin Shen}
\affiliation{Key Laboratory of Quark and Lepton Physics (MOE) and Institute of Particle Physics, Central China Normal University, Wuhan 430079, China}
\affiliation{College of Engineering and Technology, Baoshan University, Baoshan 678000, China}


\date{\today}

\begin{abstract}

Machine learning (ML) of phase transitions (PTs) has gradually become an effective approach that enables us to explore the nature of various PTs more promptly in equilibrium and nonequilibrium systems. Unlike equilibrium systems, non-equilibrium systems display more complicated and diverse features because of the extra dimension of time, which is not readily tractable, both theoretically and numerically. The combination of ML and most renowned nonequilibrium model, directed percolation (DP), led to some significant findings. In this study, ML is applied to (1+1)-d, even offspring branching annihilating random walks (BAW), whose universality class is not DP-like. The supervised learning of (1+1)-d BAW via convolutional neural networks (CNN) results in a more accurate prediction of the critical point than the Monte Carlo (MC) simulation for the same system sizes. The dynamic exponent \;$z$\; and spatial correlation length correlation exponent \;$\nu_{\perp}$\ were also measured and found to be consistent with their respective theoretical values. Furthermore, the unsupervised learning of (1+1)-d BAW via an autoencoder (AE) gives rise to a transition point, which is the same as the critical point. The latent layer of AE, through a single neuron, can be regarded as the order parameter of the system being properly re-scaled. Therefore, we believe that ML has exciting application prospects in reaction-diffusion systems such as BAW and DP.

\end{abstract}
\maketitle



\section{Introduction}
\label{intro}

Over the past decade, machine learning (ML) has shown great potential in physics research owing to its
incomparable capabilities for complex data processing and feature extraction\cite{goodfellow2016deep}. 
Various concepts regarding ML have been developed to solve vexing problems in high- energy physics~\cite{Ma2022AJT,2019A}, 
astronomy~\cite{huerta2019enabling,Emran2023PlutosSM}, quantum information~\cite{kookani2023xpookynet,Zhang_2021},and molecular dynamics~\cite{Roncoroni2023UnsupervisedLO}.

In statistical physics research, ML opens a new door to study the critical behaviours of phase transitions and demonstrates its own advantages~\cite{shen2021machine} when compared to traditional methods such as conventional field theory methods~\cite{1989Quantum,2015Field} and Monte Carlo simulations~\cite{1986Monte,1989One}. 

As the two main paradigms of ML, supervised  and unsupervised learning have shown good performance in the study of equilibrium ~\cite{christensen2005complexity,hu2017discovering} and nonequilibrium ~\cite{henkel2008non,shen2022transfer} phase transition models. Supervised learning is mainly used to identify or classify the phases of matter~\cite{2019A,Shen2021SupervisedAU}, where the input data must be labeled. Unsupervised learning does not require input labels and is more suitable for clustering and reduction of dimensions~\cite{wang2016auto,zhang2021prototypical,Emran2023PlutosSM}. However, the transfer learning method updates the algorithm structure in the existing experience and generalizes the scope of application of the model\cite{Chen2022StudyOP}.

The field of supervised learning on phase transitions and critical phenomena includes has been extensively studied. Key references include but are not limited to :~\cite{2016Learning,2019Machine,2016Detection,2020Machine,yau2022generalizability,2021A,tan2020comprehensive,2018Applications}. For unsupervised learning, the applicability of ML methods can be found in. ~\cite{2017Unsupervised,2017Machine,giataganas2022neural}. Unsupervised learning involves the use of various algorithms. Many algorithms have been applied to the study of phase transitions, such as principal component analysis (PCA)\cite{2016Discovering,2017Unsupervised}, t-distributed stochastic neighbor embedding (T-SNE)\cite{2017Unsupervised,2014Accelerating,2016How}, and nonlinear autoencoders (AE) \cite{2017Unsupervised,2017Unsupervised1}.  Additionally, the comprehensive utilization of supervised and unsupervised learning performs well for different types of models~\cite{2017Machine,2016Learning,2016How,Shen2021SupervisedAU}.

Inspired by these studies, we now focus on the branching–annihilating random walk (BAW) process, which was originally proposed in the 1980s by Grassberger \textit{et al.  } \cite{grassberger1981phase}. The BAW describes systems composed of particles that can diffuse, branch, and annihilate in a $d$ - dimensional space at different rates \cite{henkel2008non,1999A,cardy1998field,Täuber_2005,janssen2005field}.
Owing to the competition between pair annihilation and branching, a non–equilibrium phase transition occurs between the active and absorbing phases with a steady-state density of zero.
It is worth mentioning that the critical behavior of BAW is dependent on the existence of a symmetry concerning the parity of offspring number\cite{PhysRevLett.68.3060,PhysRevE.50.3623,PhysRevE.52.5955}. 
Without symmetry, the BAW system exhibits a second-order phase transition which belongs to the DP universality class. By maintaining symmetry, phase transition can be classified into the parity-conserving (PC) universality class. Owing to the rich critical phenomena of the BAW, it is a useful model for the study of non-equilibrium phase transitions in statistical physics~\cite{henkel2008non}

Considering the successful application of machine learning (ML) to phase transitions within the DP universality class, it seems logical to extend these ML techniques to other universality classes. Particularly, those that involve particle reaction-diffusion processes in their systems. This is an important motivation for our study involving the BAW process. In this study, we first attempted to measure the critical point of the BAW process using Monte Carlo (MC) techniques, which resulted in uncertainty. With smaller system sizes, we used supervised and unsupervised learning methods to identify the critical points of the two-offspring BAW processes. We determined that for small-sized BAW process clusters, convolutional neural networks (CNN) can extract basic information in the zone near the critical point and predict the critical point of the system using its forward and reverse outputs. We also measured the critical exponents using the outputs of the CNN for different system sizes and evolutionary time steps. To demonstrate the application of the unsupervised learning method, we extracted the two-dimensional structural information of the system using a basic AE to downscale the original data and analyzed the representational significance of the one-dimensional encoded output of the AE. According to our results, it is clear that for BAW processes that do not belong to the DP universality class, ML techniques are still valid, at least in one dimension. A simple CNN can still obtain basic critical-state information in small systems. Furthermore, AE, a category of unsupervised learning, can be used to extract valid information for such reaction-diffusion systems.

The remainder of this paper is organized as follows. In Sec.\uppercase\expandafter{\romannumeral2}.A, we introduce the specific definition of the BAW model in particle reaction-diffusion systems, and Sec. \uppercase\expandafter{\romannumeral2}.B is the reference for the time-dependent simulation approach for even offspring BAW. Sec.\uppercase\expandafter{\romannumeral3} is part of the MC results. Sec.\uppercase\expandafter{\romannumeral4}.A provides a specific introduction to the CNN used. In Sec.\uppercase\expandafter{\romannumeral4}.B, we provide the critical point, critical exponents of the spatial correlation length, and the characteristic time in the two-offspring BAW model. Sec.\uppercase\expandafter{\romannumeral5}.A introduces the AE in the unsupervised learning method. First, the critical point is predicted using the one-dimensional characteristic outputs of AE in Sec. \uppercase\expandafter{\romannumeral5}.B. We then explore the significance of the one-dimensional encoded outputs of AE using shuffled clusters. Finally, Sec.\uppercase\expandafter{\romannumeral6} provides the summary of this paper.

\section{Model}
\subsection{The Model of BAW}

\begin{figure*}[t]
    \centering
    \includegraphics[width=0.7\textwidth]{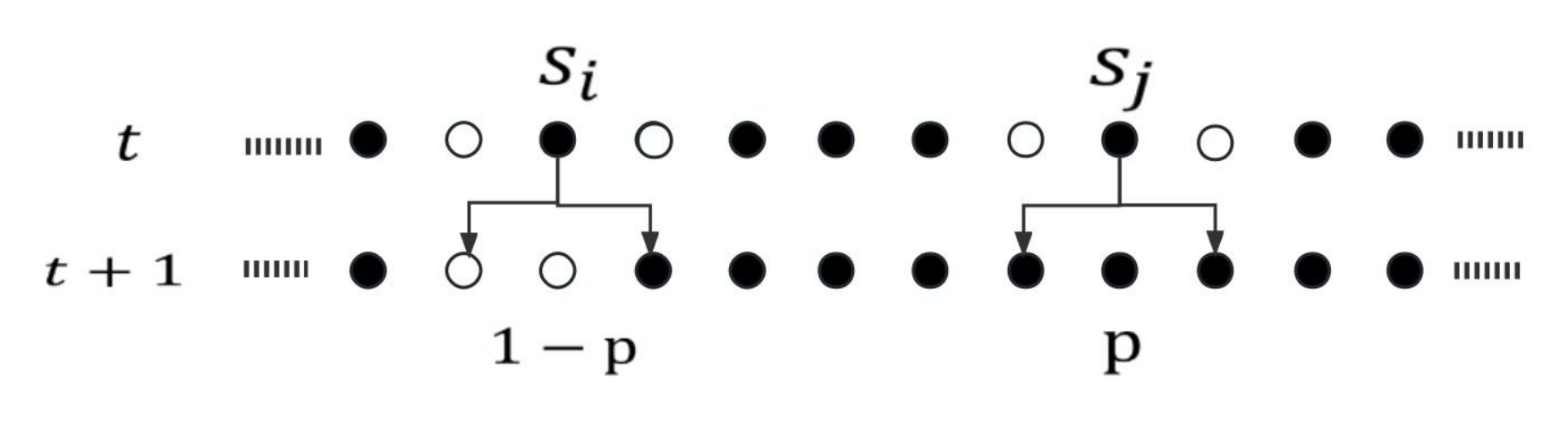} 
        
    
\caption{Nearest-neighbor occupation and the evolution results of the particles in the (1+1)-dimensional two-offspring BAW model, where the white dots represent the unoccupied particles and black ones, the occupied particles.}
\label{baw_1}
\end{figure*}

Currently, there is a lack of n ormative and generally valid formal theories for nonequilibrium systems. In the practice of dealing with the dynamic evolution of nonequilibrium systems, many theoretical methods, including the master equation of the probability distribution, mean-field approximation, and renormalization group, have been developed \cite{tauber2014critical,1989Quantum,2015Field}. In addition to theoretical attempts, numerical simulation methods, such as MC simulations, have also been applied \cite{1986Monte}. For the BAW process, we can retrieve the relevant background in the linear response of equilibrium systems \cite{1989One}. For the stochastic dynamics of particle reaction-diffusion systems, non-equilibrium systems are constructed based on Markovian processes by considering the long-term evolution forming Markov chains. Thus, the BAW processes can be related to the kinetic Ising model. One-dimensional Ising spins can be mapped to interacting particle dynamics, where particles represent domain walls on the spin chain, annihilation reactions originate from spin flips, and branching processes originate from fundamental spin exchange.

A BAW with a constant diffusion rate $D$ can be interpreted as a reaction-diffusion process of identical interacting particles $A$ and empty sites $\varnothing$ ~\cite{grassberger1981phase}: 
\begin{equation}
\begin{array}{rcl}
 \mbox{branching:} &   A \rightarrow (m+1)A, \\ 
 \mbox{annihilation:} & 2A\rightarrow \varnothing. \\
\end{array} 
\label{BAW-diff}
\end{equation} 
where $m$ denotes an integer ($m \geq 0$). The phase transition of the BAW belongs to either the DP universality class when $m$ is odd or to the PC universality class~\cite{hinrichsen2000non}. In one dimension, the dynamics of a probabilistic cellular automaton model \cite{1999A},
a kinetic Ising model \cite{1989One,1995Non},
and an interacting monomer-dimer model \cite{Menyh_rd_1996,1996Reaction}
is equivalent to the class of BAW.
For $m = 2$, the one-dimensional BAW model can characterize dynamic Ising models, which violate a detailed balance~\cite{1999A,1989Some}. Particle A represents the domain walls, and the transition to the inactive state corresponds to the ordering of Ising spins~\cite{Balboni_1995}.
Common methods for studying BAW include mean-field \cite{cardy1996theory}, master equation \cite{cardy1998field},  and Monte Carlo simulations \cite{zhong1995universality}.
For a one-dimensional BAW, we applied the ML approach to study the case of $m=1$ in ~\cite{shen2022transfer,Shen2021SupervisedAU}, termed as $(1+1)$-dimensional DP model. We now shift our focus to the case of even numbers of $m$, termed as BAW with $m$-offspring which belongs to the PC universality class.

To define BAW in one-dimensional lattices, a particle is first randomly selected at time $t$, and then it may branch to, with probability $p$, or diffuse to, with probability $1-p$~\cite{zhong1995universality}, a randomly chosen nearest neighbor at time $t+1$. The particle jumps to a randomly chosen nearest neighbor with probability $1-p$, and the incident and target particles are annihilated with probability $c$ if this site is already occupied. With probability $p$, the particle produces $m$ offspring at the nearest neighbor sites. Annihilation occurs at rate $c$ if offspring are created at an occupied site. Fig.\ref{baw_1} shows the evolution of \;$S_i$\; and \;$S_j$\;points at time \;$t+1$\; when diffusion and branching reactions are selected at time \;$t$\;, respectively.
For a one-dimensional BAW with an infinite annihilation rate($c=1$), regardless of the branching rate, the system always evolves into a vacuum-absorbing state when $m = 2$, resulting in no phase transition\cite{2016A}. To examine the transition behavior for $m = 2$, a finite reaction rate must be introduced\cite{1995Non,1994Propagation}, which is set to $c=1/2$ \cite{1999Conservation,PhysRevE.50.3623,zhong1995universality}.

\begin{figure}
\centering
\includegraphics[width=0.28\textwidth]{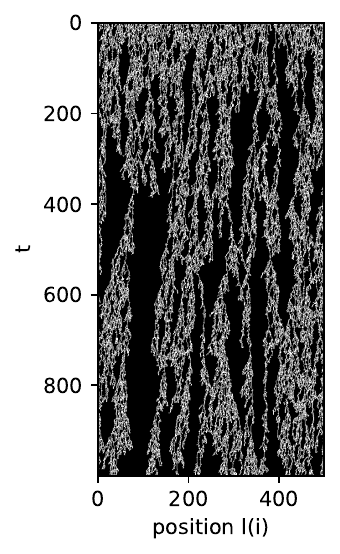}
\caption{Cluster diagrams of the system in the homogeneous initial state of the (1+1)-dimensional BAW model, where the branching probability is $0.80$ and annihilation probability is $0.50$. System size is set to \;$L=500$\;, \;$t= 1000$\;.}
\label{cluster1}
\end{figure}

To study critical behavior, the evolution of the system can be described by the concept of ``average time". With the number of currently surviving particles $N$, the average time is set and counted as $\Delta t=1/N$ when a particle is selected at each time; Hence, in a time step, each particle either branches or diffuses once on average. As the annihilation rate $c$ is fixed, the dynamics of the system can be evaluated by probability $p$ only with time step $t$.

\begin{figure*}[t]
    \centering
    \includegraphics[width=0.65\textwidth]{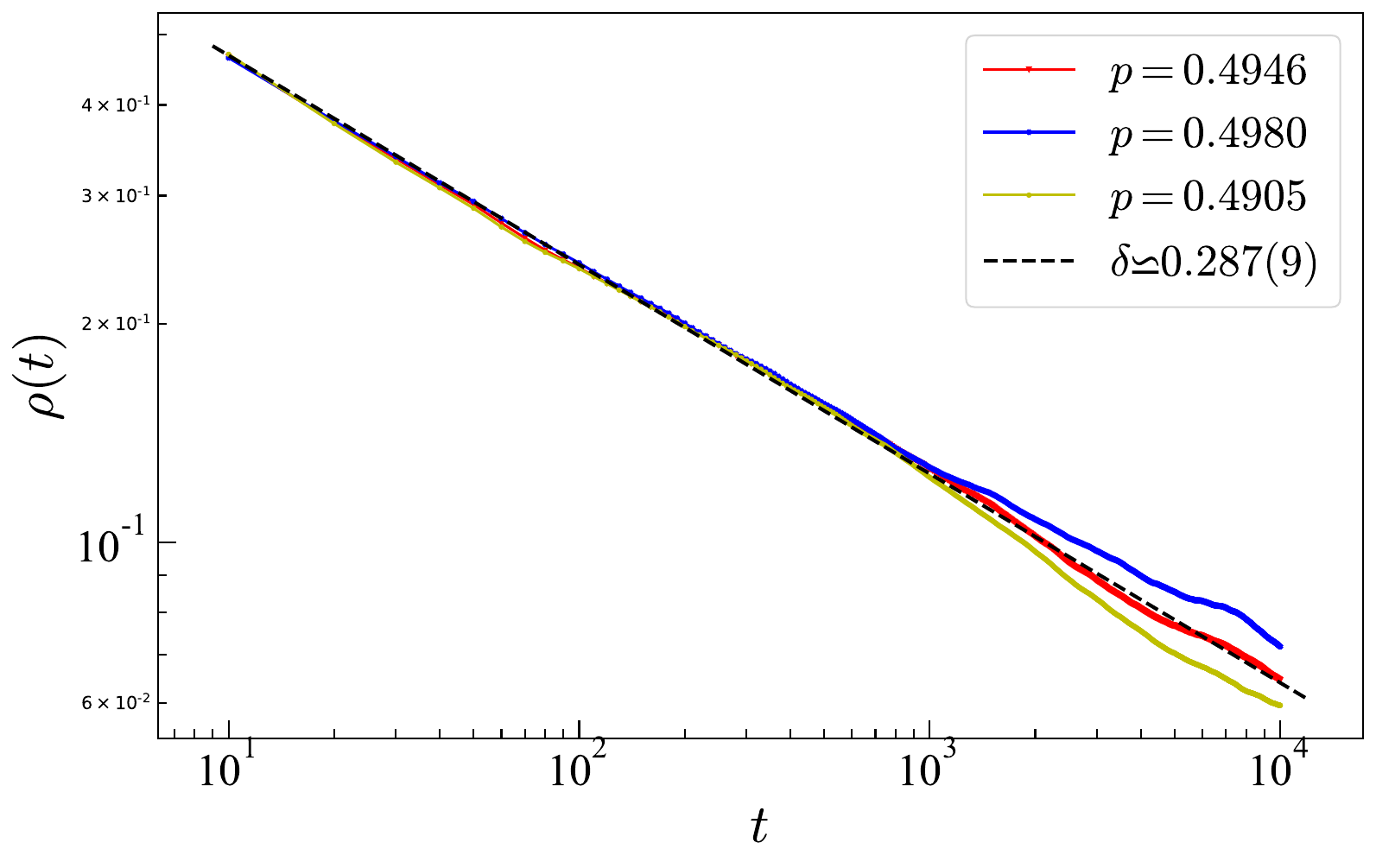} 
\caption{MC simulations of (1+1)-dimensional two-offspring BAW model, where the lattice size is \;$L = 15000$\;, total time step is \;$t = 10000$\;, and the number of ensemble average is \;$5000$\; for different values of \;$p$\; corresponding to \;$0.4980$\; (blue), \;$0.4946$\; (red), and \;$0.4905 $\;(yellow). The black dashed line represents a power-law fitting of the data points in red, with a slope of $ -\delta \backsimeq -0.287(9)$.}
\label{pofbaw}
\end{figure*}

\subsection{Time-Dependent MC Simulations}

In this study, time-dependent simulations of the (1+1)-dimensional BAW process are implemented using the MC method following Ref. ~\cite{zhong1995universality}.
We commence from the configurations close to the absorbing state on the grid, and the system evolves with the ``average" time $\Delta t$ as independent implementations. 
With the time evolution, the critical behavior of the branching rate $p\to p_c$ can be measured by the quantities governed by the power laws with $t\to \infty$:
\begin{equation}
P(t) \sim t^{-\delta}, \quad \langle n(t) \rangle\sim t^{\eta},\quad \langle R^2 (t)\rangle\sim t^{z},
\label{delta}
\end{equation}
where $P(t)$ denotes the probability that the system has not entered the absorbing state at $t$, $\langle n(t) \rangle$ denotes the mean number of particles at time $t$ averaged over all runs, $\langle R^2 (t)\rangle$ denotes the mean square distance of spreading from the centre of the lattice averaged over only the surviving runs.

Another important critical exponent is $\beta$, which is related to the stationary concentration of particles in steady state~\cite{henkel2008non}:
\begin{equation}
\rho_a \sim (p-p_{c})^{\beta}, 
\end{equation}
where $\rho_a$ denotes the particle density. The steady state of BAW follows the power law:
\begin{equation}
\xi_{\perp}\sim (p-p_{c})^{-\nu_{\perp}},\quad \xi_{\parallel}\sim (p-p_{c})^{-\nu_\parallel},
\label{critical_exp}
\end{equation}
where $\xi_{\perp}$ denotes the spatial correlation length, $\nu_{\perp}$ denotes the spatial correlation exponent, $\xi_{\parallel}$ denotes the temporal correlation length, and $\nu_\parallel$ denotes the temporal correlation exponent. 

Determining the exact values of a series of critical exponents is important for determining the universality of a model. The lattice evolution model, which is designed based on the BAW process, may differ in the measurement of the critical point. However, the measurement of the critical exponent of the PC universal class should be consistent. Given the requirements of the theoretical methods for the thermodynamic limit, we used cyclic boundary conditions in the numerical simulation to reduce the finite-size effect. The accuracy obtained by measuring the critical points in a finite lattice is limited because of computational limitations. We elaborate on this in Sec. \uppercase\expandafter{\romannumeral3}.

\section{MC of two-offspring BAW}

Based on the evolutionary processes described in Sec. \uppercase\expandafter{\romannumeral2}, a BAW with a two-offspring process evolution program is designed. Based on this, the evolution cluster diagrams are extracted. In Fig. \ref{cluster1}, we show the cluster evolution structure of the BAW model with system size \;$L=500$\; and time step \;$t = 1000$\; under high branching probability and finite annihilation rate. Long-range correlations and cluster characteristics of the system are observed. In contrast to the DP model, the absorbing state of the system appears under the condition of an even seed and a sufficiently long evolution time step based on the dynamic exponent\;$z$\;.

Simulations with the above sizes are certainly insufficient to approach the critical behavior of the system under the ideal condition of an infinite size. Under the condition of rapidity-reversal symmetry, the critical exponent of the particle survival rate can be determined using the time-dependent decay form of the statistical particle density\cite{henkel2008non}. We consider the attenuation form of the system particle density at a critical point.
\begin{equation}
\varrho(t) {\quad}{\sim}{\quad} t^{-\delta}.
\label{delta}
\end{equation}

Under the settings of size \;$L=15000$\; and time step \;$t=10000$\; with homogeneous initial states, we calculated the change in the particle density of the system with time. Fig. \ref{pofbaw} shows the change in system's particle density under different branching probabilities and the power-law fitting result. To determine the power-law decay interval of the particle densities, we measured the goodness of fit for three different branching probabilities. 
\begin{equation}
R^2=1-\frac{\sum\left(y_a-y_p\right)^2}{\sum\left(y_a-y_m\right)^2},
\end{equation}
where \;$y_a$\; denotes the actual value of the density of the statistics, \;$y_p$\; denotes the corresponding predicted value of the fitting, and \;$y_m$\; denotes the average of the actual values. When branching probabilities \;${p}$\; are 0.4946, 0.4980, and 0.4905, the goodness of fit of the particle density to the decaying power law, \;$R^2$\;, is 0.9982, 0.9484, and 0.9553, respectively. Based on the fact that the optimal value of \;$R^2$\; is \;$1$\;, the power-law decay of particle density performs better when the branching probability is $0.4946$.

As shown in Fig. \ref{pofbaw}, although the particle density shows a power-law like attenuation at the branching probability \;$p=0.4946$\;, we don't have a good deviation at \;$p=0.4980$\; and \;$p=0.4905$\;. Therefore, we can only ensure that the critical point of the system is within the range \;(0.4905,0.4980)\;. From the diagrams, the results of using the MC technology to analyse the critical point of the system will cause some uncertainty. The attenuation of the system particle density below the critical point was not as obvious. Moreover, for system sizes of L=15000 and t=10000, the computational force requirement is relatively demanding. Working with smaller sizes, ML can obtain more accurate critical point estimations using small-sized system cluster diagrams. For the other critical exponents of the PC universality class, we turn to ML.

\begin{figure*}[t]
    \centering
        \includegraphics[width=1\textwidth]{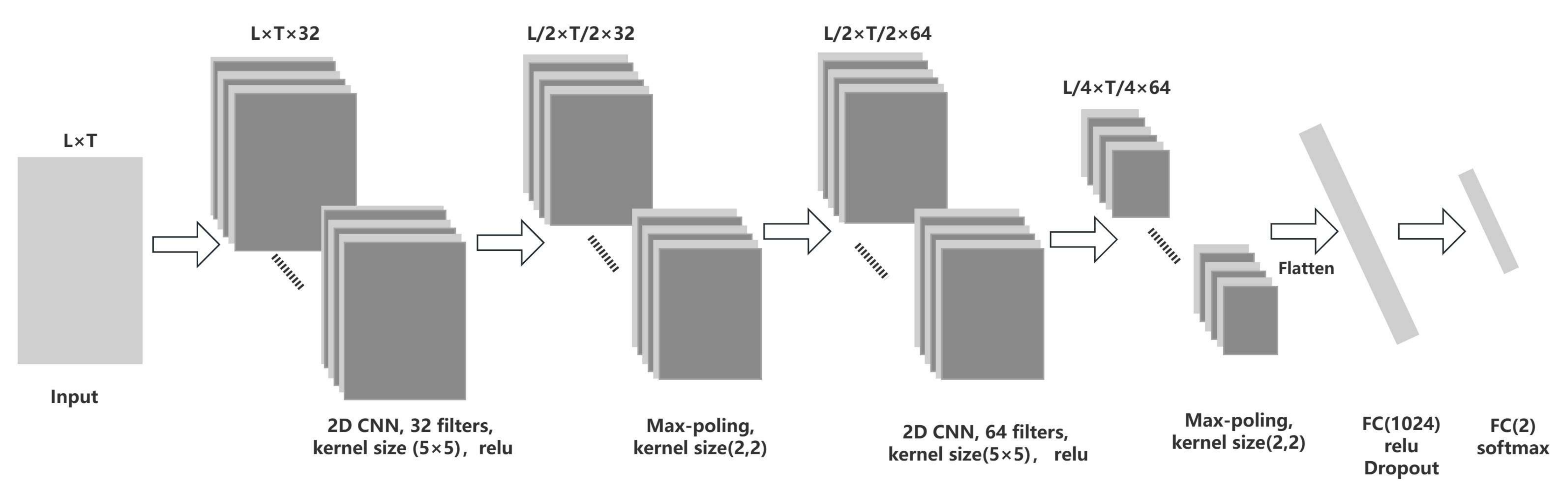} 
\caption{General structure of CNN.}
\label{cnnstru}
\end{figure*}

\begin{figure}[!htb]
\includegraphics[width=0.45\textwidth]{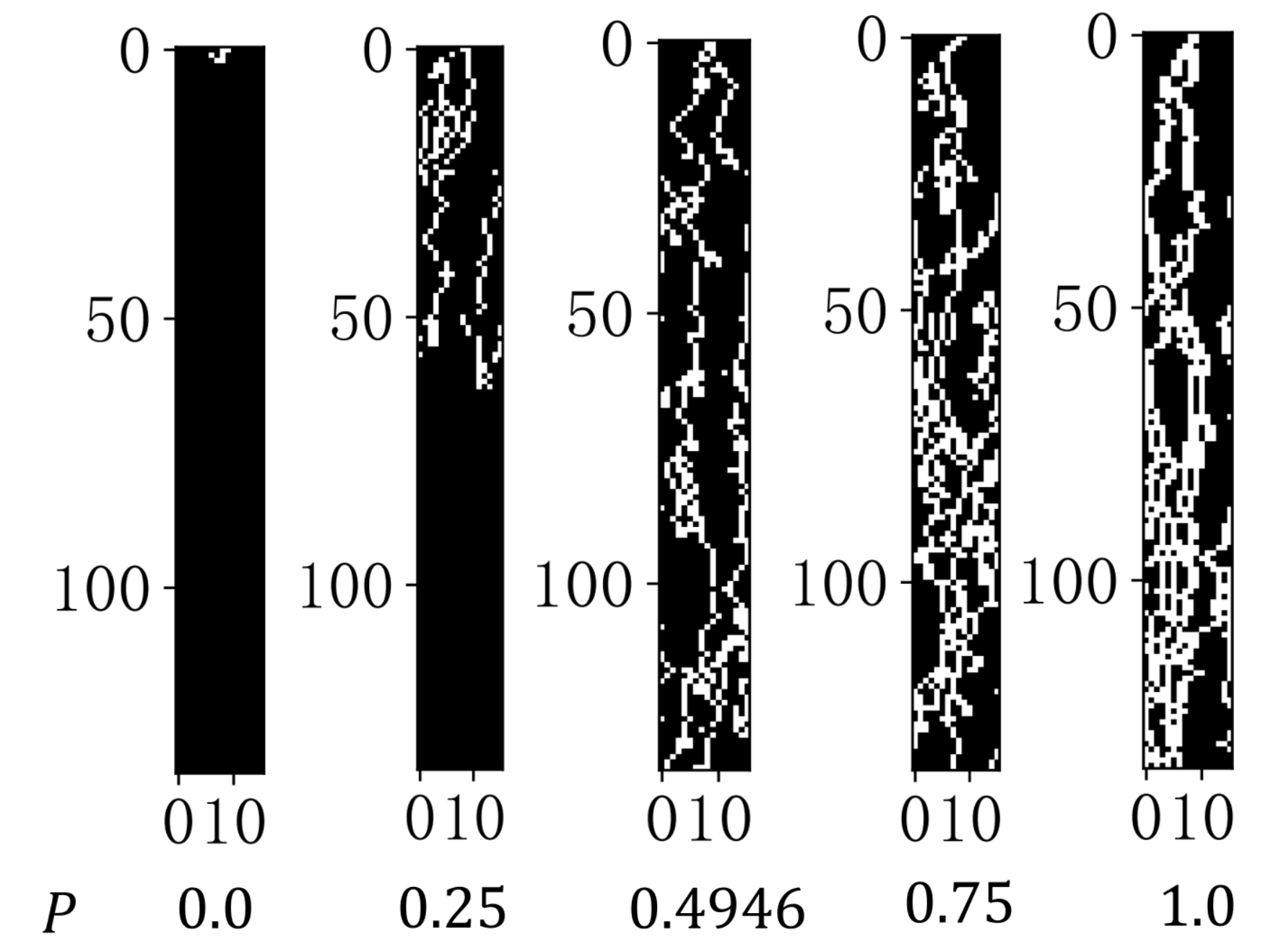}

\caption{Evolution results of representative cluster diagrams controlled by different branching probabilities of the BAW model under pair-seed initial conditions, with branching probabilities from left to right, are $p=0, 0.25, 0.4946, 0.75, 1.0$ and the system size is $L = 16$.}
\label{cluster2}
\end{figure}

\begin{figure*}[t]
\begin{tabular}{cc}
    \includegraphics[width=0.46\textwidth]{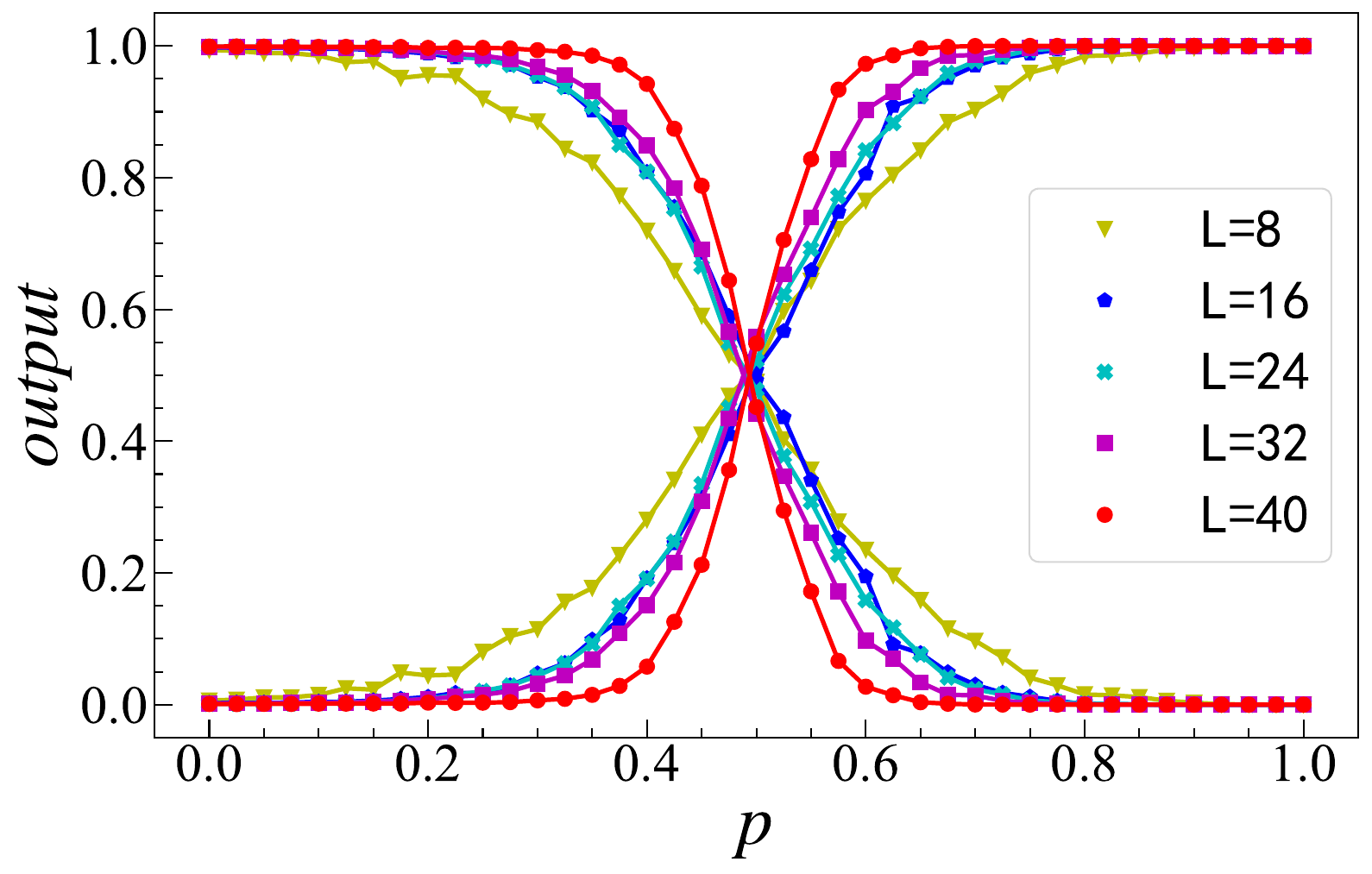} &
    $\qquad$\includegraphics[width=0.46\textwidth]{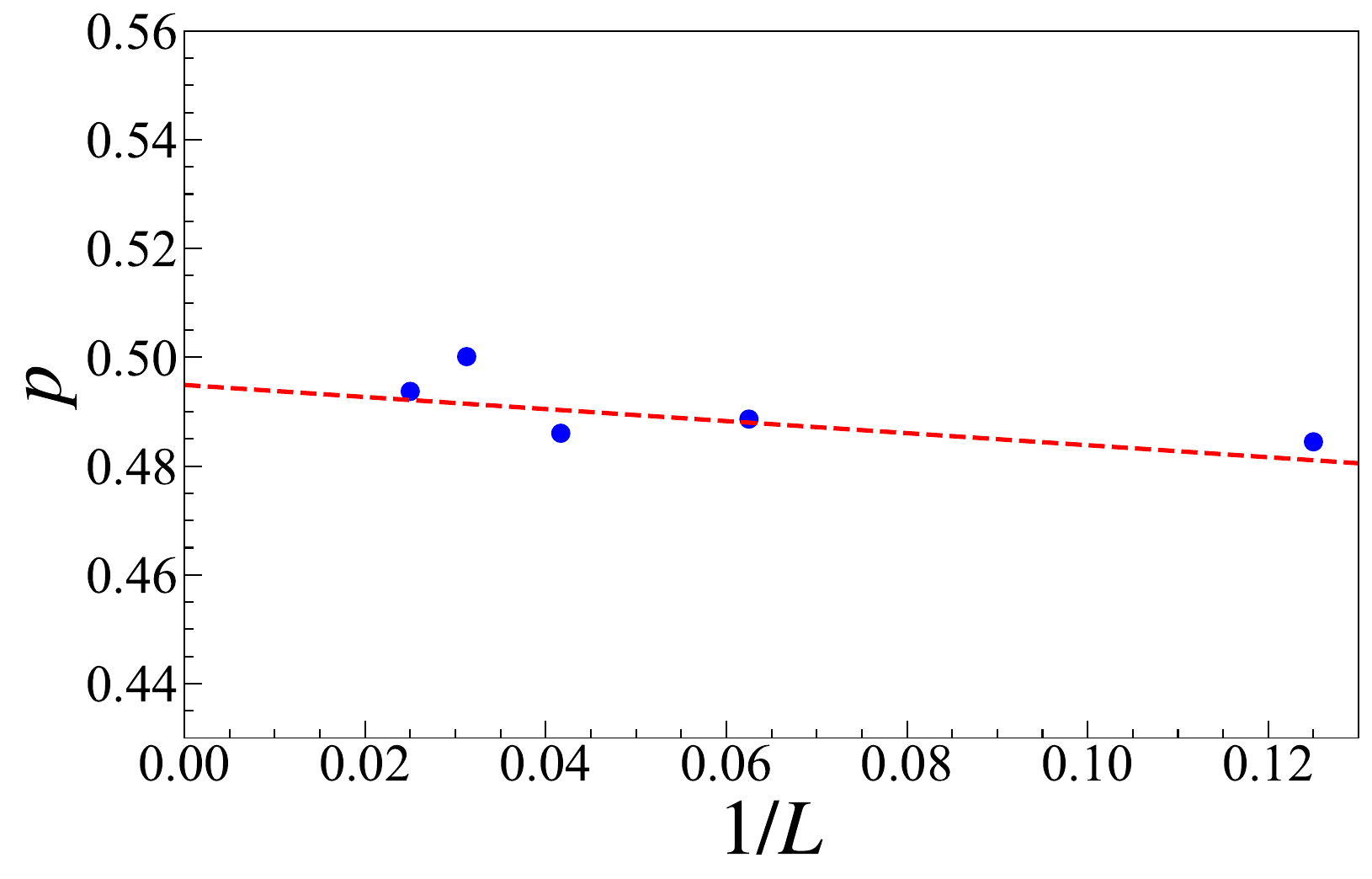}  \\
    {\quad}{\quad}(a) & $\qquad$ {\quad}{\quad}(b)
\end{tabular}
\caption{(a)Outputs of CNN at different sizes \;$L = 8,16,24,36,40$\;. (b)CNN with different system sizes identifies the critical point, in which the polynomial fitting result is \;$-0.1107x+0.4949$\;, indicating that the predicted value of the critical point of the system at infinite size is \;$p_c=0.494(9)$\;.}
\label{841t}
\end{figure*}

\begin{figure}[ht]
\centering
\includegraphics[width=0.45\textwidth]{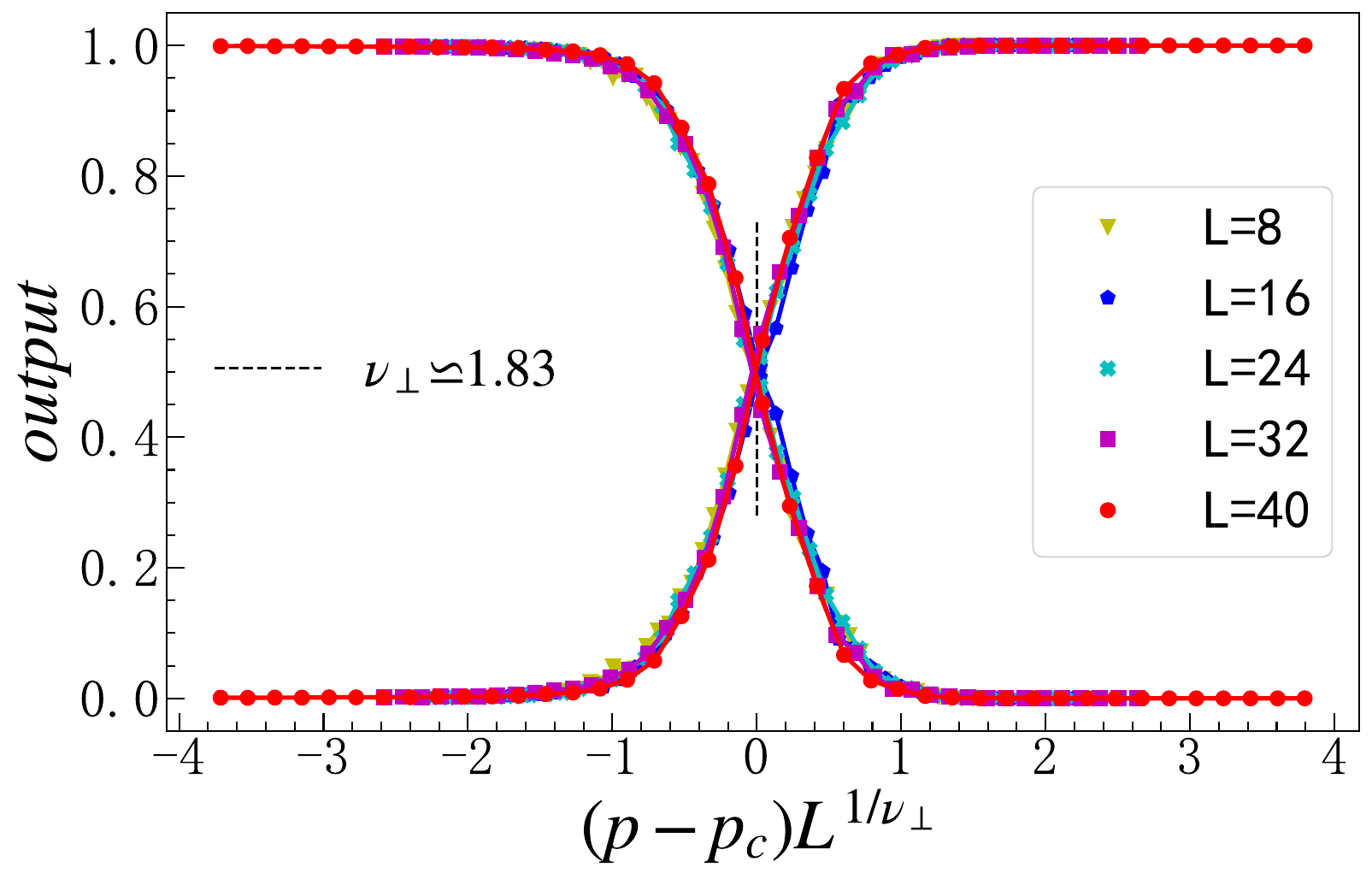}
\caption{Data collapse of the CNN outputs in different sizes after using finite-size scaling law, which yields \;$\nu_\perp=1.83$\;.}
\label{nu_prep}
\end{figure}

\begin{figure*}[t]
\begin{tabular}{cc}
    \includegraphics[width=0.46\textwidth]{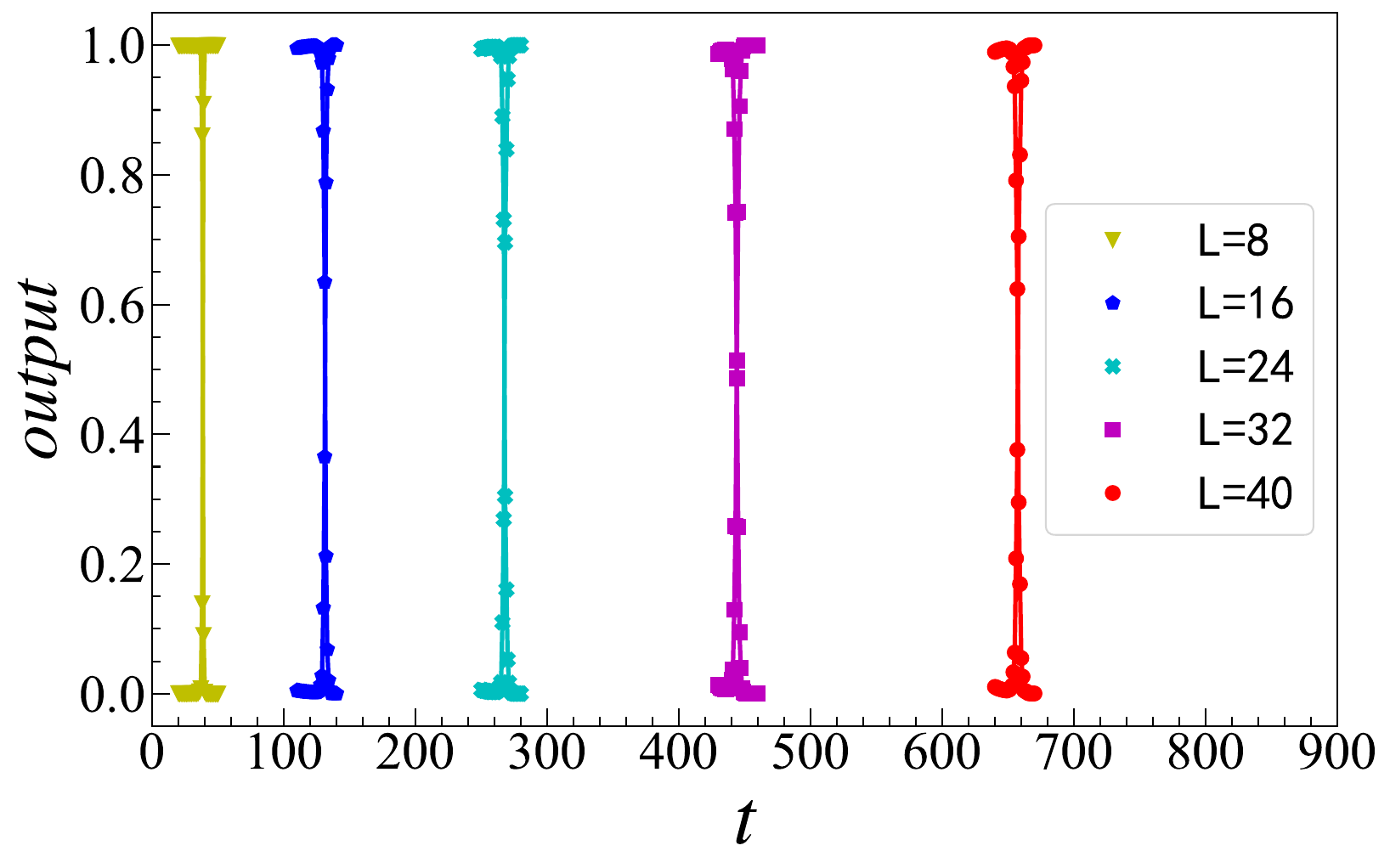} &
    $\qquad$\includegraphics[width=0.47\textwidth]{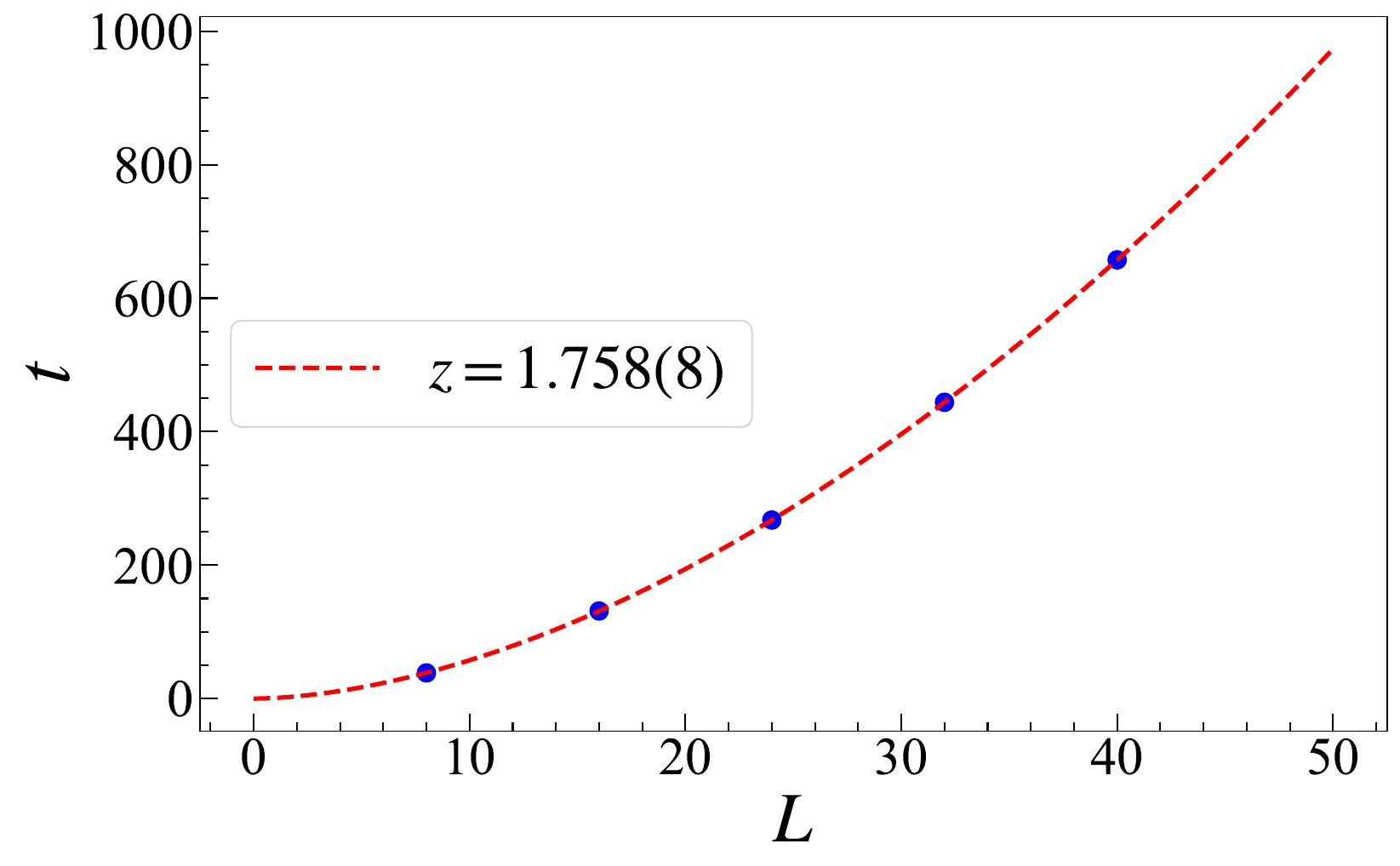} \\
    {\quad}{\quad}(a) & $\qquad$ {\quad}{\quad}(b)
\end{tabular}
\caption{(a)Uniform display of CNN outputs at different system sizes. See text for specific sizes and time steps selection. With a fixed control parameter ($p=0.4949$), the intersections of CNN outputs predict the position of the characteristic time steps. (b) Recognition of characteristic time steps by CNN with different system sizes. Synthesizing the training results of CNN under different sizes \;$L=8, 16, 24, 32, 40$\;, we record the predictive values of characteristic time and perform an exponential fitting, which yields the estimated value of the dynamic exponent, \;$z=1.758(8)$\;.}
\label{1t}
\end{figure*}

\section{Supervised learning of BAW}
\subsection{Method of CNN}
As a popular approach for regression and classification~\cite{goodfellow2016deep,huerta2019enabling,2019A}, the CNN has proven to be a powerful tool for dealing with the critical behaviors of different phase transition models ~\cite{shen2021machine,Shen2021SupervisedAU}.
Typically, the architecture of a CNN comprises convolution, pooling, and classification layers. There can be more than one convolution and pooling layer, if needed. The detailed structure of a CNN is shown in Fig~\ref{cnnstru}.

First, the clusters of a finite-size system at a specific evolution time step are received as inputs by the CNN. The system exhibits absorbing and active phases under different branching probabilities. For this purpose, a CNN is employed to identify the phase features of the input data, and its structure, which is fixed at the beginning, remains the same for all clusters with different branching probabilities. To facilitate phase classification, we use a convolution layer of $32$ filters with $5\times 5$ kernel size and maximum pooling layer with $2\times 2$ kernel, followed by another convolution layer of $64$ combined with maximum pooling to extract the features of the configurations more accurately. To combine these features with a wider variety of attributes, they are flattened and then fed to a fully connected layer with $1024$ neurons such that NN has a better capacity to map the objective function. The RELU activation function of form $f(x)=max(0,x)$ provides a nonlinear mapping. We also introduce a dropout layer in the network structure to lower the training costs and prevent overfitting. Finally, the $softmax$ function assigns a probability to each category, which could be an outcome of the classification.

For the training process, parameter $p$ is assigned $41$ different values from range $[0,1]$, and 2000 samples of configurations are generated for each value.
For the training process, we label the data according to the results of MC in Section \uppercase\expandafter{\romannumeral3}: the configurations with branching probabilities below $0.4946$ are assigned '$0$, and those with branching probabilities higher than $0.4946$ are assigned '$1$'. To converge the model, 10 epochs are used for each training set. Conversely, $1/10$ of the data of the training set without labels are chosen as test data. The basic idea is to use a well-trained network to predict the critical point using the classification power of a CNN. The loss function chosen is the mean square error (MSE). We use Python 3.7 and TensorFlow 2.3.0.

\subsection{Learning result of CNN}

To achieve stable results for generating configurations with MC, the maximum time step is slightly higher than the characteristic time $t_c$, proportional to $L^{z/d}$, where $z = 1.75(9)$ and dimension $d=1$~\cite{henkel2008non}. In the study of (1+1)-dimensional BAW,  we use a two-dimensional CNN fed by the input of configurations, as shown in Fig~\ref{cluster2} with lattice length $L=16$ and time step $T=135$. We use system sizes of $L=8,16,24,32$, and $40$, and the configurations shown in Fig. ~\ref{cluster2} provide examples of inputs to the supervised and unsupervised ML learning in the following sections. Under the setting of branching probabilities less than $0.4946$, we generate training data with label '$0$'. When the branching probabilities are higher than $0.4946$, training data labeled as '$1$’ are generated. The relationship $t_c{\sim}L^z$ between the dynamic exponent $z$ and characteristic time $t_c$ indicates the general criterion for the absorbing and active states of the system at a specific size. This involves controlling the size of the branching rate to detect the presence of active particles at time step $t_c$. Given that the test procedure is unlabelled, we hypothesize that the CNN captures this evolutionary feature.

The CNN test results of (1+1)-dimensional two-offspring BAW at $L=8,16,24,32$ and $40$ are shown in Fig. ~\ref{841t}. This is illustrated using a pair of red curves for the two output predictions of CNN at $L = 40$. When $p$ moves from 0 to 1, the prediction probability of the configurations in the active phase increases and that in the absorbing phase decreases. Under the aforementioned labeling method, the intersection point of the CNN output is very close to the critical point \cite{2017Machine,2019Machine}. 

According to finite-size scaling (FSS) theory \cite{2000Theory,PhysRevLett.28.1516,1990Finite}, the critical point corresponding to an infinite lattice size, $p_c$, can be extracted by extrapolating $1/L$ to zero with linear fitting. To obtain a more accurate measurement of $p_c$, we perform five independent runs of the CNN for the same dataset at each $L$. After fitting, the critical value of the prediction $p_c$ at $L\to\infty$ is $0.494(9)$, which is consistent with the MC simulation of 0.4946 \cite{zhong1995universality}. For a relatively small system size, the MC cannot obtain an accurate position of the critical point. A CNN can reduce the computational cost of determining the critical point and providing reference intervals with higher accuracy.

The results of the CNN shown in Fig. ~\ref{841t} can be fitted with a sigmoid function of scaling width $\sigma$
\begin{equation}
 p \rightarrow 1-\frac1{1+e^{\frac{-(p-p_c)}{\sigma}}} \,,\frac1{1+e^{\frac{-(p-p_c)}{\sigma}}}
 \label{eq:sigmoid}
\end{equation}
Specifically, it is straightforward to measure the spatial correlation exponent $\nu_{\perp}$ using a data collapse process. Using the sigmoid function parameters, the output predictions are plotted as functions of $(p-p_c)/\sigma$. According to FSS in the correlation length ${\xi}_{\perp}$
\begin{equation}
{\xi}_{\perp}\sim L \sim|p-p_c|^{-\nu_{\perp}}{\rightarrow{|p-p_c|}}\sim L^{-1/\nu_{\perp}},
\end{equation}
we can obtain the scaling of the form $(p-p_c )L^{1/\nu_{\perp}}$, as shown in Fig~\ref{nu_prep}. We maintain the vertical-axis data unchanged and change the horizontal-axis coordinates to $(p-p_c )L^{1/\nu_{\perp}}$. Specifically, $p_c$ is fixed at $0.4949$, and the effect of data collapse is realized by adjusting $\nu_{\perp}$ to $1.83$ in Fig. ~\ref{nu_prep}. By fitting, we obtain a value of $\nu_{\perp}\simeq 1.83$, which is close to the MC value of $1.83(3)$.

However, the derivation of the power exponent $z$ makes our prediction more convincing. Considering the dynamic evolution of the system, we identify the characteristic time $t_c\sim L^z$ of the system near the critical point $0.494(9)$ and use different time steps to generate configurations~\cite{Shen2021SupervisedAU}. 

For the CNN training process for different $L$, we select $1000$ samples for each time step. Clusters whose time steps are less than the characteristic time $t_c$ are labeled as $0$, whereas those with time steps greater than $t_c$ are labeled as $1$. For each $L$ only $31$ time steps adjacent to $t_c$ are selected to avoid enlarging the data. Fig.~\ref{1t}(a) shows the output of CNN with $t=[20,50],[110,140],[250,280],[430,460],[640,670]$ at $L=8, 16, 24, 32, 40$, and Fig.~\ref{1t}(b) presents the power-law fitting of each intersection in Fig. ~\ref{1t}(a) with $5$ independent runs. Based on this, $z \simeq 1.758(8)$ is close to the MC value $z=1.75(9)$.

\begin{figure*}[t]
        \centering
        \includegraphics[width=0.83\textwidth]{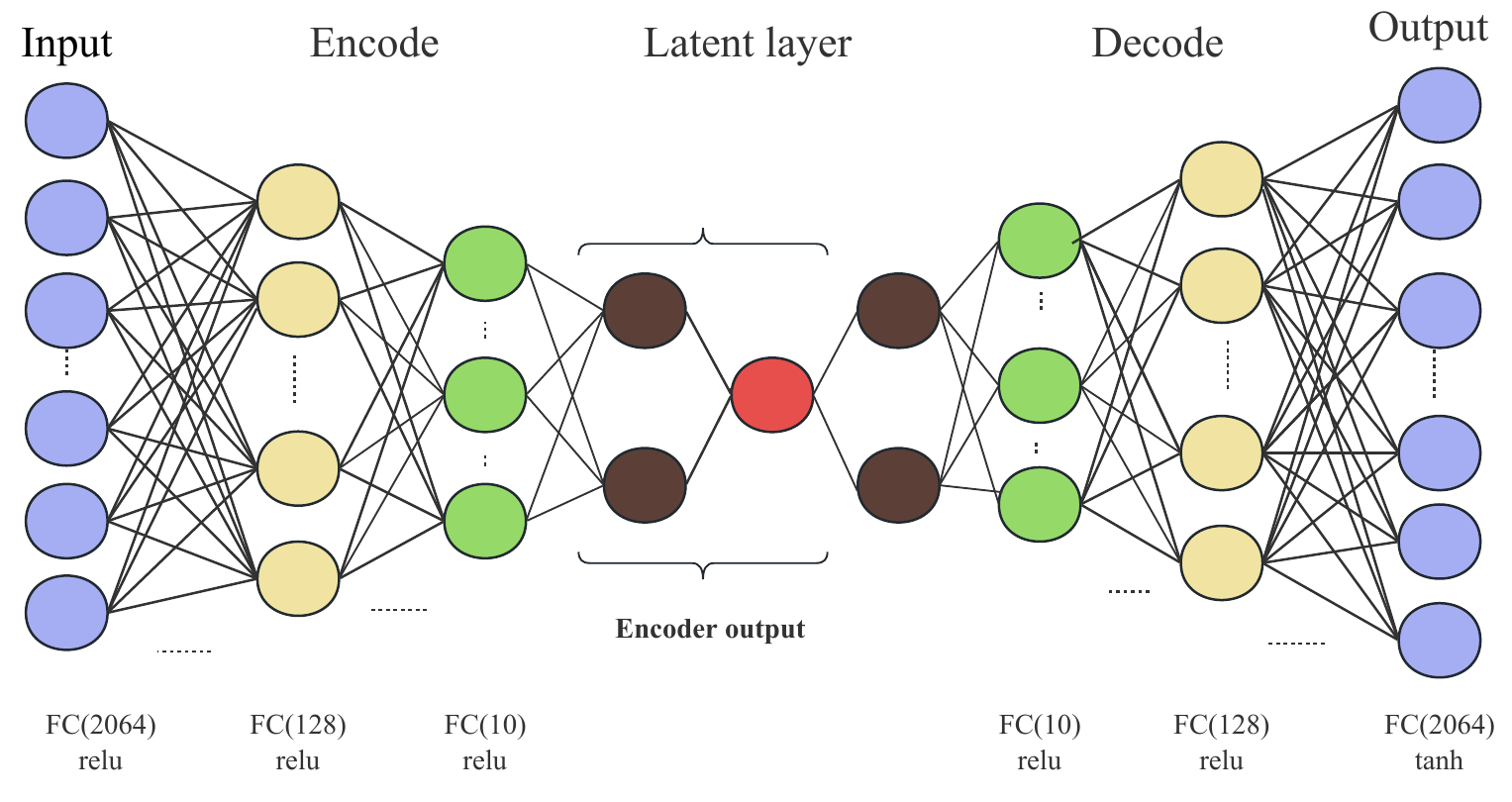} 
        \centerline{}
\caption{Neural network schematic structure of AE. The “Encoder output” part indicated by the brackets in the figure contains the encoded results. The output can be either two-dimensional with two neurons, colored in brown, or one-dimensional with one neuron, colored in red.}
\label{AEstruc}
\end{figure*}

\begin{figure}
\centering
\includegraphics[width=0.50\textwidth]{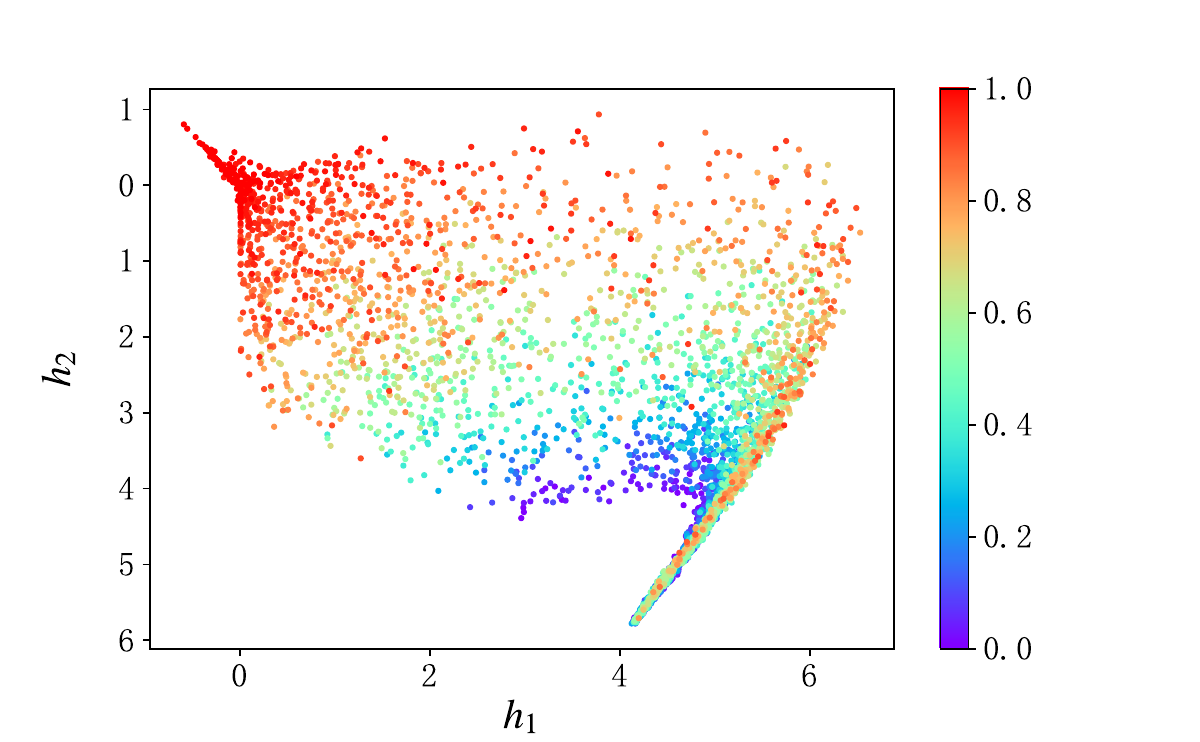}
\caption{Two-dimensional feature extraction of (1+1)-dimensional BAW model encoding by AE. The branching probabilities are represented by different colors of the color bars beside. When approaching the critical point, the points show the characteristics of fuzzy dispersion.}
\label{range1}
\end{figure}

\begin{figure*}[t]
\begin{tabular}{cc}
    \includegraphics[width=0.43\textwidth]{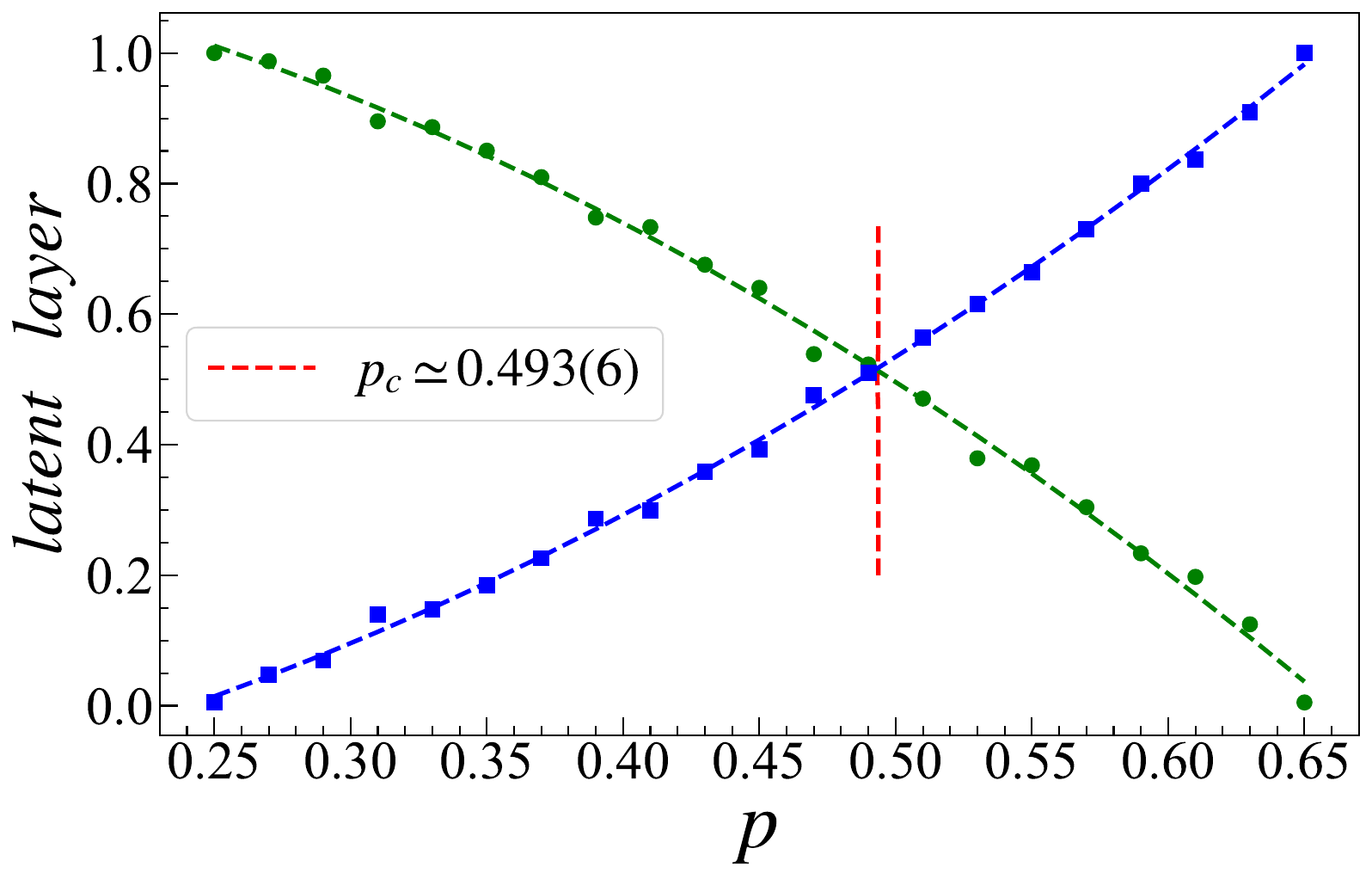} &
    $\qquad$\includegraphics[width=0.43\textwidth]{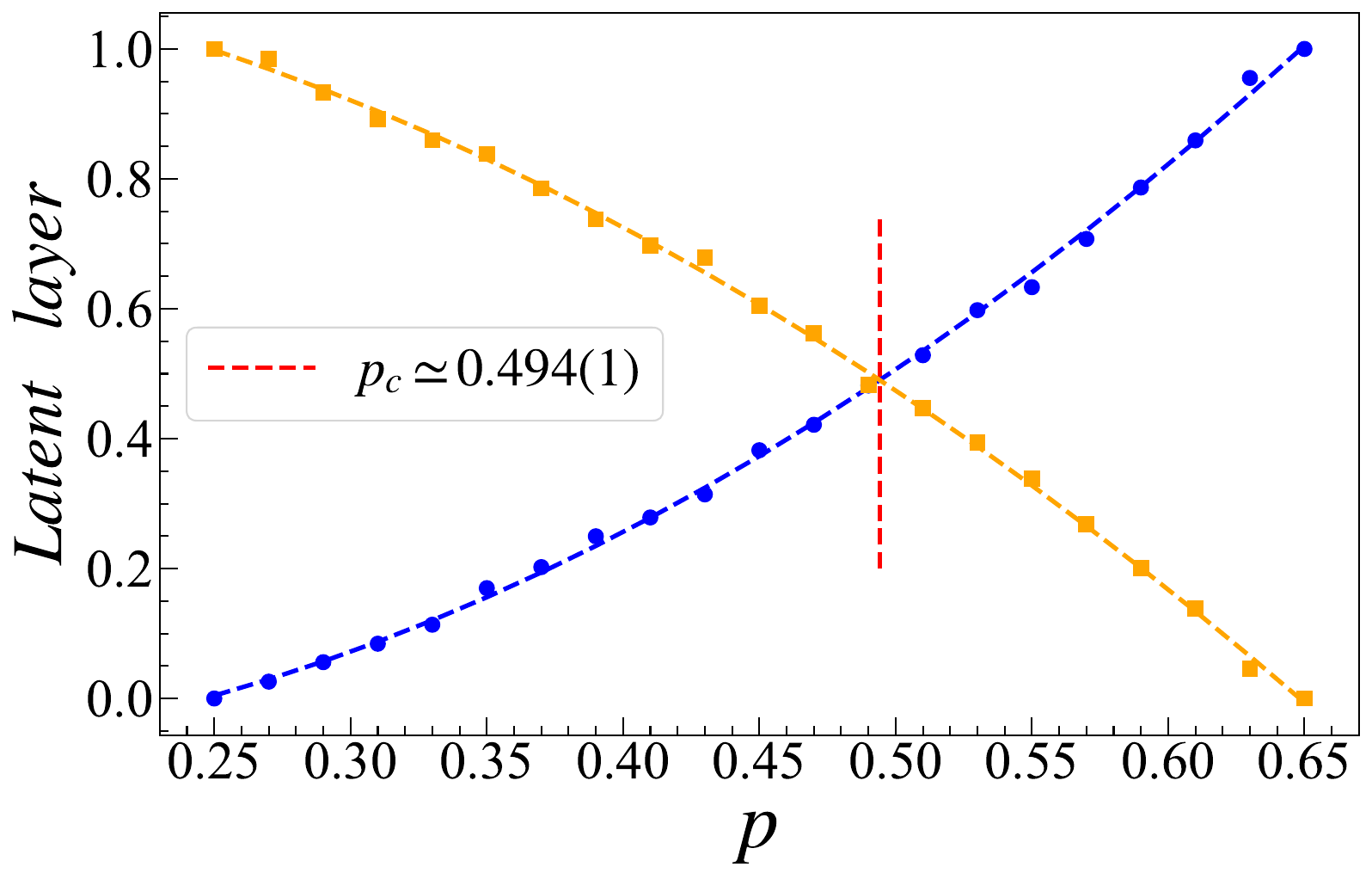} \\
    {\quad}{\quad}(a) & $\qquad$ {\quad}{\quad}(b)
\end{tabular}
\caption{(a)One-dimensional normalized outputs of AE with two different trends when the system size is \;$L=16$\;. The green and blue curves are the polynomial fitting results of the basic AE outputs. The ensemble average is considered to reduce the statistical error. The intersection point is located at p=0.493(6). (b) One-dimensional outputs of AE when \;$L=40$\;. The intersection point is located at p=0.494(1).}
\label{9t1}
\end{figure*}

\begin{figure*}[t]
\begin{tabular}{cc}
    \includegraphics[width=0.43\textwidth]{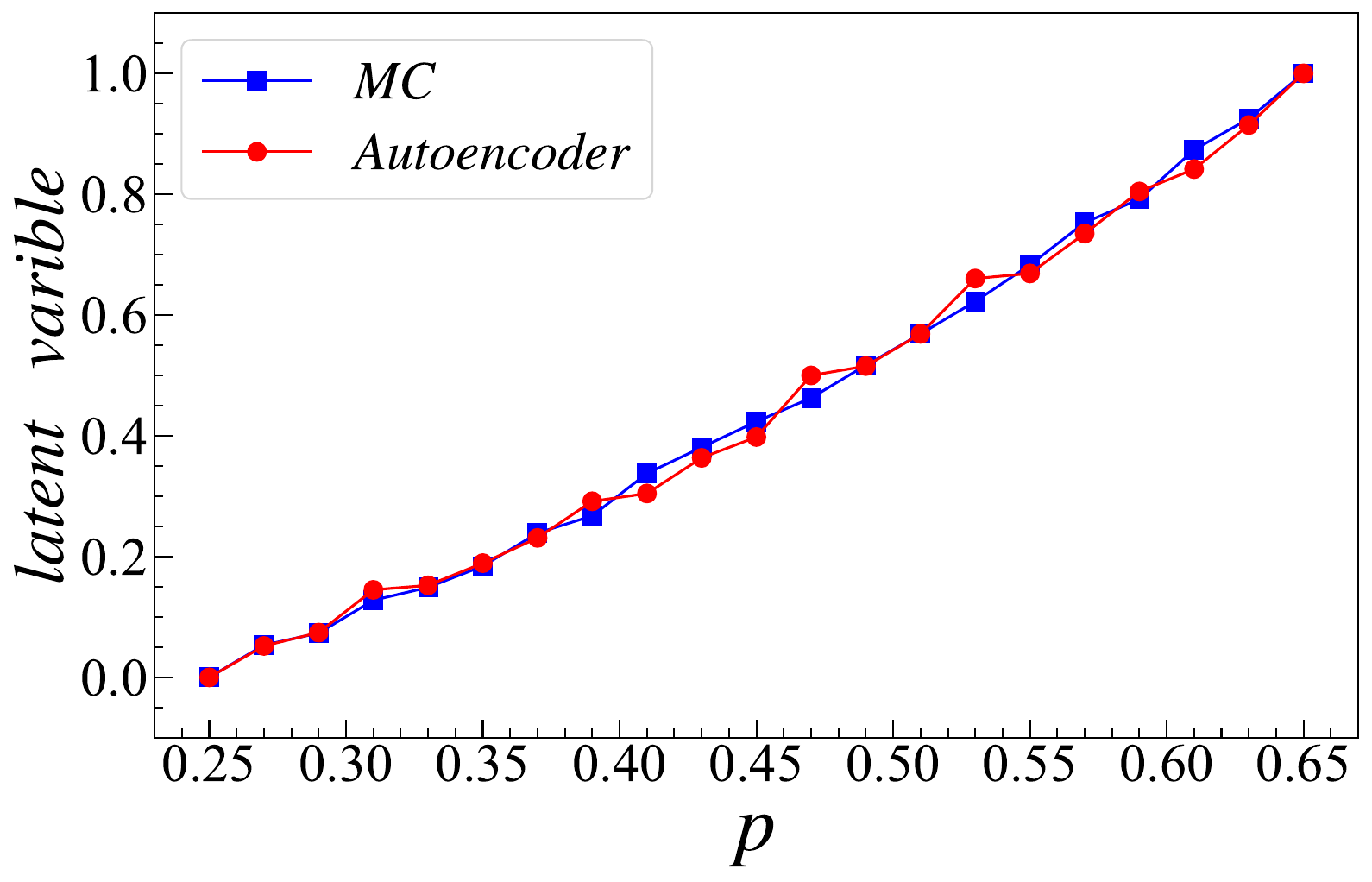} &
    $\qquad$\includegraphics[width=0.43\textwidth]{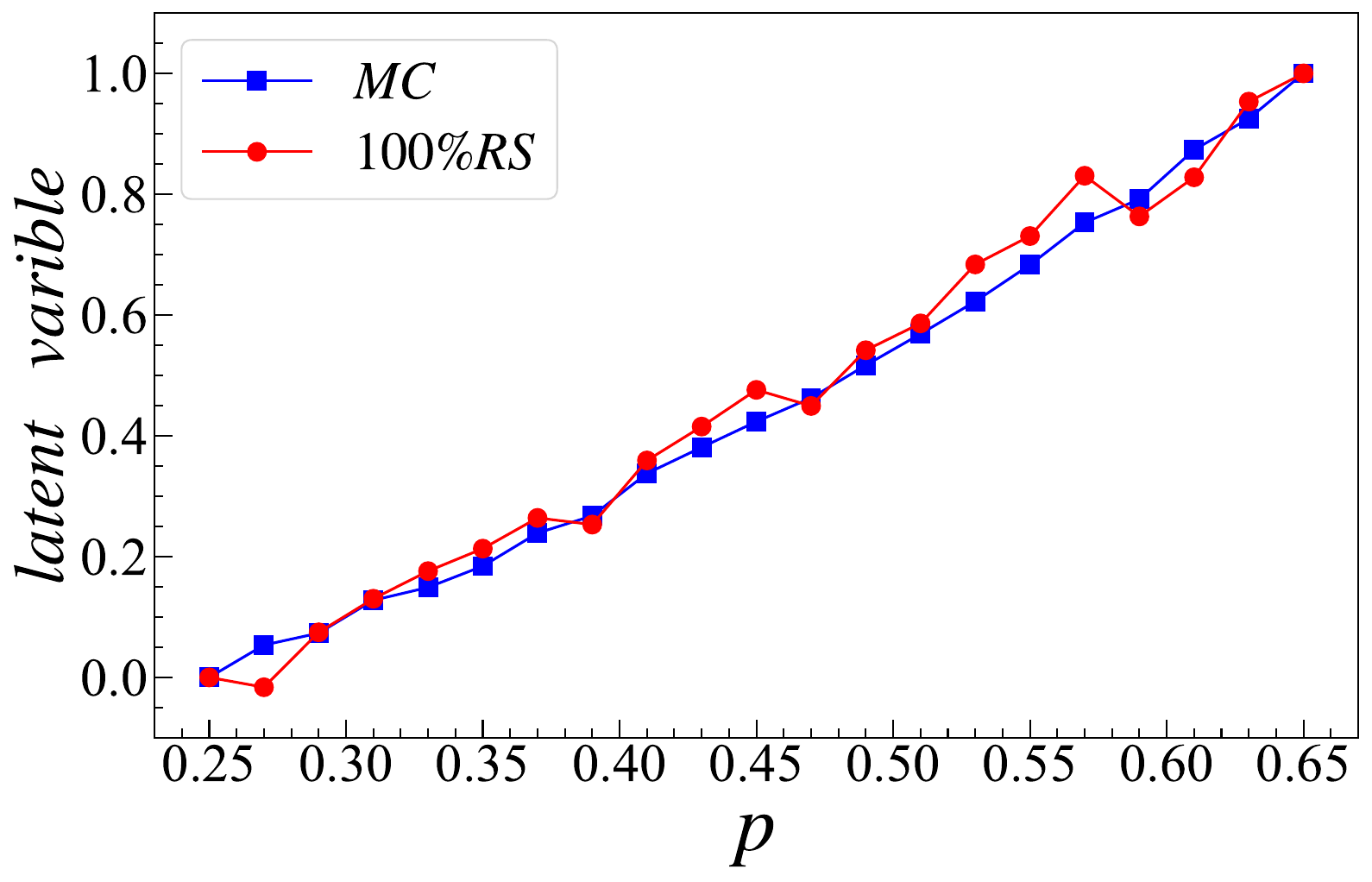} \\
    {\quad}{\quad}(a) & $\qquad$ {\quad}{\quad}(b)
\end{tabular}
\caption{(a)Statistical comparison of the one-dimensional output of AE and system's particle density when \;$L=16$\;. The blue points are the normalized result of the particle density, and the red points are the normalized result of the AE's one-dimensional output. It can be observed that the two plots almost collapse, and the Pearson correlation coefficient between them is 0.9979. (b) The statistical comparison between the particle density and one-dimensional output of AE after the clusters are completely shuffled. The Euclidean distance between the two plots is 0.0945, which indicates that they almost collapse.}
\label{9t2}
\end{figure*}

\begin{figure}
\centering
\includegraphics[width=0.18\textwidth]{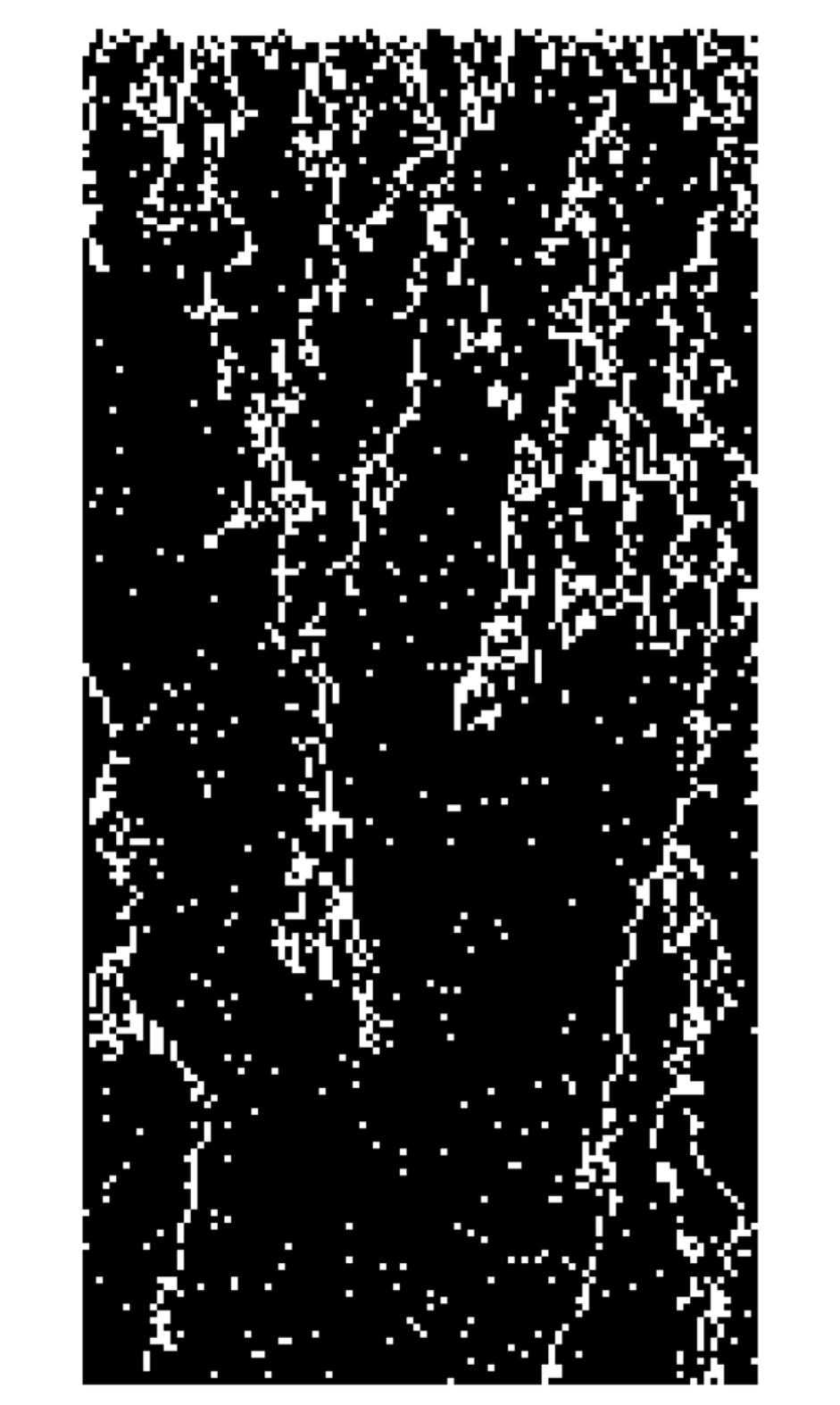}
\caption{Structure of the cluster with size $L=100,T=200$ after \;$20\%$\; of sites are shuffled. The effect of partial disruption is similar to system noise.}  
\label{range}
\end{figure}

\section{Unsupervised learning of BAW}
\subsection{Approach of the autoencoder}

Unlike supervised learning, unsupervised learning algorithms automatically classify or group input data with reference to unlabeled training data and process the data according to their own structural characteristics~\cite{goodfellow2016deep}. 
It is well known that an AE is a type of NN used to reconstruct images, text, etc., from compressed versions.
The basic mechanism of autoencoding is that it learns the implicit features of the input data through an encoder and reconstructs the original input from the newly learned features through a decoder \cite{1991Nonlinear}. 
Given their ability to extract hidden information, AE have been used to deal with phase transitions with promising results~\cite{Shen2021SupervisedAU,shen2021machine}. In this study, we applied the AE to the $(1+1)$-dimensional even-offspring BAW to explore its ability to learn critical behaviors and extract the critical features of the model.

The structure of an AE based on an FCN is illustrated in Fig. ~\ref{AEstruc}. The input data are the same as those used for the supervised learning, except that the training data are unlabeled. The pool of configurations is divided into three parts: $50\%$ for training, $30\%$ for validation, and $20\%$ for testing. During the encoding and decoding processes, a dynamic learning rate is adopted to optimise the parameter configuration. Adam~\cite{2014Adam} is used to introduce a momentum correction bias for the parameter updates. A regularization step is added to reduce the noise in the training data. Cross-entropy is selected as the loss function to evaluate the reconstruction ability.

The structure of the AE demonstrates its capability for data reconstruction and dimension reduction. Simultaneously, the encoding process ensures that the data features can be extracted effectively, which can then be used in the deconstruction of the phase diagram information. We use the AE's data dimensionality reduction function to detect the critical points of the BAW model. Therefore, in this study, we focus on the encoding process of the AE.

The basic workflow of AE is as follows: first, the cluster diagrams with different branching probabilities are inputted into the AE, corresponding to the "Input" layer in Fig. ~\ref{AEstruc}. As shown on the left side of Fig. ~\ref{AEstruc}, the AE then encodes the raw data through the fully connected layers by progressively decreasing the number of neurons and applies the RELU activation function to create a nonlinear mapping. After encoding, the dimensions of the original input data are reduced, and the new dimensionality is determined by specifying the number of neurons in the "encoder output" layer. Decoding is the reverse process of encoding, which aims to utilize the encoded low-dimensional results to reshape the original cluster diagrams. During multiple back propagations with parameter updates, cross-entropy is used as the loss function to evaluate the effectiveness of the data reconstruction. After completing AE training, the encoded results with different numbers of neurons are obtained. The results for the two neurons are indicated in brown in Fig. ~\ref{AEstruc}, and the encoder output of one neuron is colored in red. For example in Fig.~\ref{range1}, the results for the two neurons are indicated by different colors. The encoder output of one neuron is represented by a specific color, which is shown in red in Fig.\ref{9t2}(a).

During training, the decoding process is the reverse of the encoding process. The decoding output is backpropagated and mapped to high-dimensional data. To improve the decoding accuracy for data dimensionality reduction, the hyperparameters of NN are updated through multiple cycles. As a test, the NN is limited to two hidden neurons, $(h_1, h_2)$. The extracted features are shown in Fig. ~\ref{range1}. The color bar displaying the different branching probabilities is shown on the right side of Fig. ~\ref{range1}. The clusters of features from the input configuration within $[0,1]$ branching probabilities can be distinguished by color, and it is evident that the features after the extraction process still reflect the characteristics of the original data. Based on the clustering of the data points in Fig. 10, it can be observed that the distribution of the data points is more spread out when approaching a critical point. This feature may correspond to the phase characteristics near the criticality. However, this is not an observable transition phenomenon, as reported in Ref. ~\cite{Shen2021SupervisedAU,shen2021machine}, even when tested using denser configurations.

\begin{figure*}[t]
\begin{tabular}{cc}
    \includegraphics[width=0.45\textwidth]{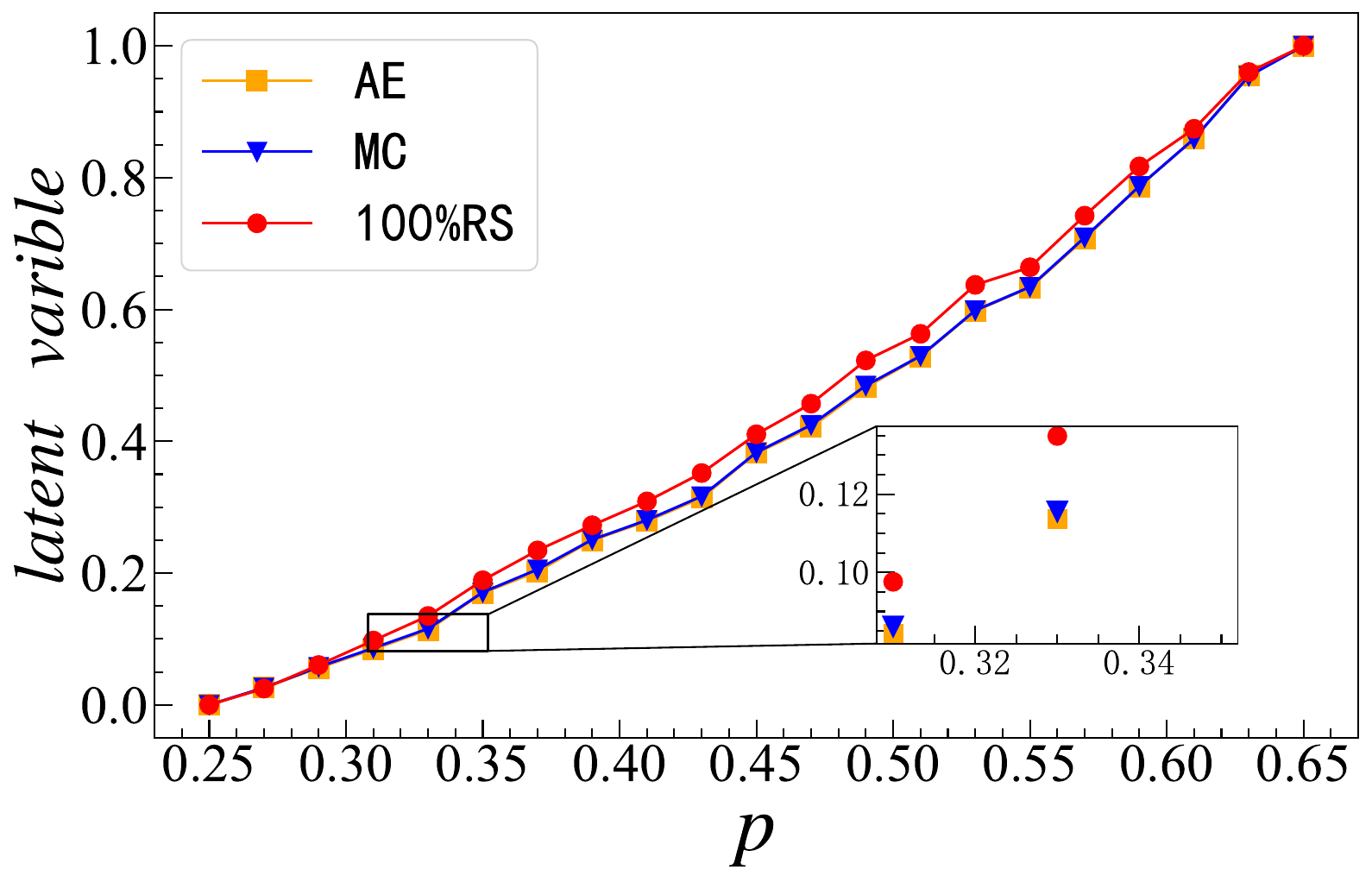}  & 
    $\qquad$\includegraphics[width=0.45\textwidth]{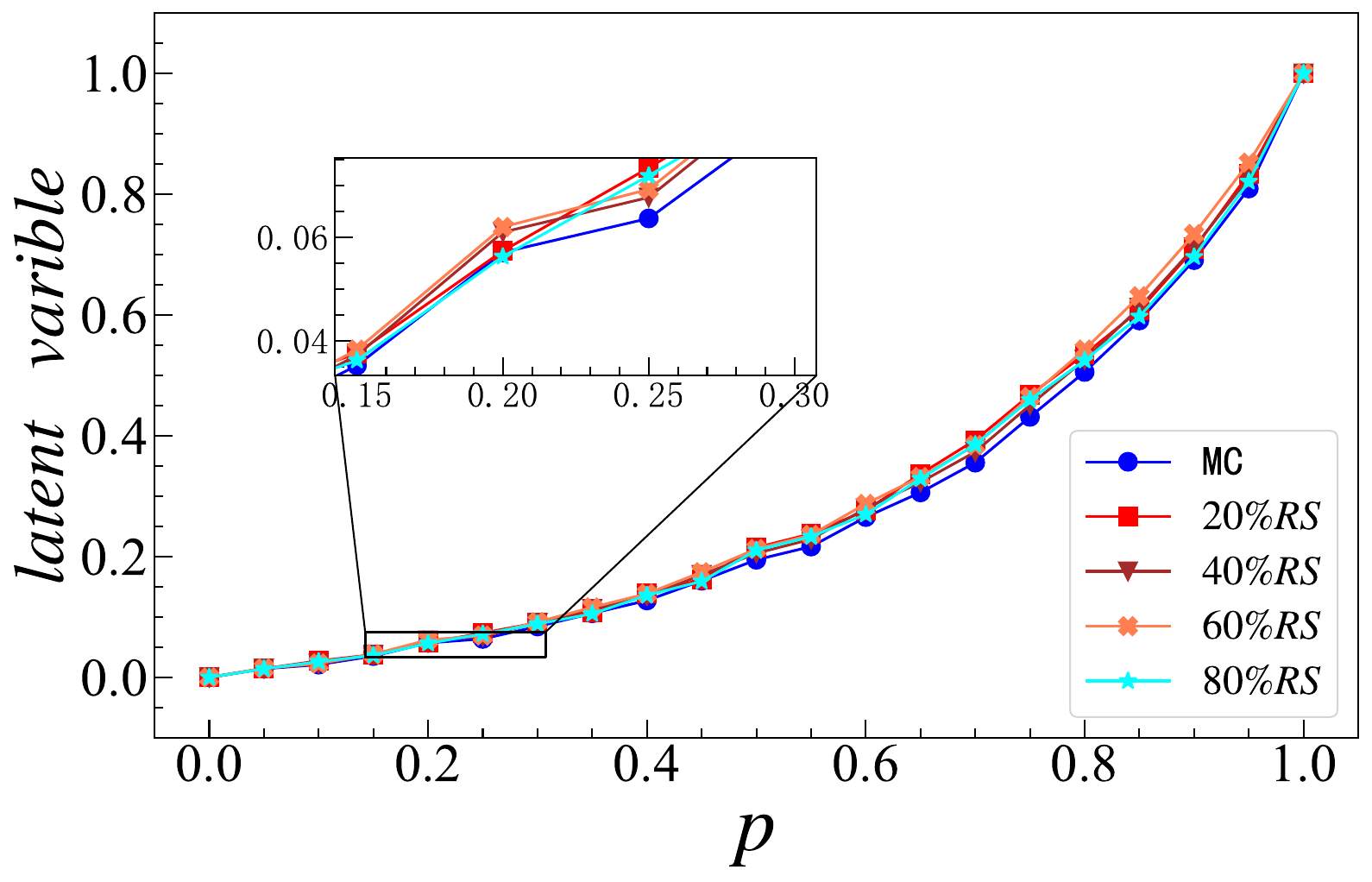} \\
    {\quad}{\quad}(a) & $\qquad$ {\quad}{\quad}(b)
\end{tabular}
\caption{(a)Statistical comparison of AE output, particle density, and AE output with completely shuffled clusters when \;$L=40$\;. (b)Comparison between AE's outputs with different shuffle rates and particle densities. The Euclidean distances between them are listed in Table \ref{table3}.}
\label{9t3}
\end{figure*}

\subsection{Learning result of the autoencoder}

By studying the characteristics of two-dimensional encoded data, we conjecture that an AE can learn the phase characteristics of clusters under different branching probabilities. However, the two-dimensional encoded results of the AE do not accurately reflect the location of the critical point. Therefore, we switch to other methods, which may explicitly represent the information of interest. A good candidate is a single latent variable in which the information of the order parameters (or equivalently, the particle density for some models) is hidden. Therefore, a one-dimensional representation of the latent variables is provided to locate the critical point. We use clusters with different branching probabilities, similar to those in Fig. ~\ref{cluster2}, as the training and test data, which are received by the AE as the input. After repeating the training of the AE several times, we observe two different trends of one-dimensional encoded results, as shown in Fig. ~\ref{9t1}.

From the viewpoint of ML, learning the structural features of BAW model clusters via AE provides the output with different dimensional features. The AE's output intersection implies an overlap of features, indicating the specificity of the clusters near the critical point. Inspired by our study on supervised learning, the intersection of downscaled feature output by the AE can characterize the location of critical points. Therefore, the two AE's outputs are normalised, and their intersection provides the location of the critical point. To obtain more information near the critical point, we narrow the selection interval of the branching probabilities. Specifically, we test $21$ values for every interval of $0.02$ in the range of $[0.25,0.65]$. The outputs of the two cross curves through the AE at $L=16$ and $L=40$ are shown in Fig. ~\ref{9t1}. The locations of the intersections are $0.493(6)$\ and $0.494(1)$. We tentatively determine the location of the critical point characterized by branching probability $p=0.494(1)$.

\begin{table*}[!tbp]
	\centering
\resizebox{350pt}{6mm}{	
	\begin{tabular}{cccccc}
			\hline
			\textbf{Shuffling degree} & \textbf{20\%} & \textbf{40\%} & \textbf{60\%}& \textbf{80\%}& \textbf{100\%}\\
			\hline
			Euclidean distance & 0.0838 & 0.0578 & 0.1050 & 0.0590 & 0.0945\\
			\hline
		\end{tabular}
		}

\caption{Statistical comparison between AE's one-dimensional coding outputs (normalized) and the particle density under different degrees of shuffling.}
\label{table3}
\end{table*}

We are also interested in analysing the relationship between the one-dimensional output of the AE and particle density of the clusters. To realize this, the AE's one-dimensional output will be divided by its maximum value. Owing to changes in the branching probability interval, the statistical values of the particle density are normalised. Fig.\ref{9t2}(a) shows a statistical comparison of the AE's one-dimensional output with the normalized particle density of the system. We use 'MC' in Fig.\ref{9t2} to refer to the normalized particle density. It can be observed that the two curves nearly collapse onto one another, and the Pearson correlation coefficient between them is $\;0.9979$\;.

\subsection{Shuffled Clusters in the autoencoder}

Considering the similarity between the particle densities and one-dimensional encoded results of AE with different branching probabilities, we analyze the ability of AE to extract the structural features of such systems. We attempt to establish a link between the one-dimensional encoded outputs of the AE and order parameter of the system. For this purpose, we retain the particle density information of the clusters while adding noise to their spatial structure. This involves shuffling the clusters by randomly switching the positions of the occupied lattices and empty ones. Some or all lattices participate in the shuffling process. The degree of shuffling can be defined as the ratio of sites involved in shuffling. The shuffled clusters are used as inputs for learning the AE. We then use the Euclidean distance to characterize the difference between the results with the original clusters and those with shuffled clusters.
\begin{equation}
d(x, y)=\sqrt{\sum_{i=1}^n\left(x_i-y_i\right)^2}
\end{equation}

First, we shuffle the original clusters at a ratio of 1 at $L=16$ and examine the encoding results using the shuffled clusters received by the AE as the input. Fig. \ref{9t2}(b) shows the comparison between the normalized particle density and outputs of the AE. Given the similarity between the two, we examine the decoding outputs of the AE using clusters with different degrees of shuffling.
Fig.\ref{range} shows the cluster results with the degree of shuffling corresponding to\;$20\%$\;. If the degree of shuffling is low, then the shuffling operation can be viewed as noise introduced into the system. After normalizing the data, Fig.\ref{9t3} shows the comparison between the particle density and AE's one-dimensional output at different levels of shuffle.
The results for the Euclidean distances are listed in Table \ref{table3}. Based on the results, the AE extraction of the structural features of the two-offspring BAW process clusters is limited to the system order parameter \;$\rho$\; (particle density of the cluster), and there is no significant change as the degree of shuffling is gradually increased. Therefore, it can be inferred that the one-dimensional output of the AE  reflects the particle density of the system.

\section{Conclusion}

In this study, the critical behavior of (1+1)-dimensional two-offspring BAW process is examined using CNN and AE methods. This study includes the determination of the critical point and measurement of critical exponents. Finally, we propose a preliminary conjecture on the AE one-dimensional coding output. Our conclusions are as follows:


First, we use the (1+1)-dimensional two-offspring BAW model evolution rules to design a  cluster generation program and compare the structural characteristics of different clusters under odd and even seed conditions. We select specific training and testing data for supervised learning and use the CNN results to identify and predict the critical point under the infinite size limit. We obtain the result \;$p_c = 0.494(9)$\;, which is very close to the MC simulation result for large sizes. We then identify the characteristic time using a CNN to determine the dynamic exponent \;$z$\;. The dynamic exponent is identified as $z = 1.758(8)$. Furthermore, we measure the critical exponent \;$\nu_\perp$\; according to the finite-size scaling law, and the result is $\nu_\perp = 1.83$.

Unsupervised learning mainly analyzes the phase classification effect of an AE's two-dimensional encoded results and uses the one-dimensional bidirectional encoded data of the AE to identify the critical point \;$p_c = 0.494(1)$\;. We also attempt to apprehend the basic AE's one-dimensional coding output of a basic AE. In the case of a gradually increasing system noise, we determined that the AE's one-dimensional encoded results can still characterise the order parameter of the system's particle density.

For this type of reaction-diffusion systems, the system sizes and time-step requests from the MC are very large. In this case, the accuracy of the critical point estimates must be improved. Based on the perspective of theoretical reference, MC quantifies macroscopic properties such as particle density and mean square distance. However, it offers limited access to the microstructure information generated during the temporal evolution of the system. Owing to the sufficiency of training data, ML can capture more microscopic information about the system using cluster diagrams, which may explain the effectiveness of ML in accurately identifying the critical points of such systems and measuring critical exponents. We believe that many models belonging to DP and PC universality classes can be examined using ML technology, which provides a new idea for the critical state analysis of such reaction-diffusion systems.

In summary, we believe that CNN and AE can be successfully applied to the BAW model to analyse the critical behaviour of the system and predict the critical exponents. Based on the perspective of the system size, ML is expected to become a universal and convenient technology. In view of the limitations of practical computing power, ML provides a new approach for examining the critical behaviour of nonequilibrium phase transitions. Exploring the relationship between the outputs of deep neural networks and the order parameters of the system can aid in developing an application of this method in the branch of phase transitions.

\section{Acknowledgements}

This study was supported in part by the Key Laboratory of Quark and Lepton Physics (MOE), Central China Normal University(Grant No.QLPL2022P01), the National Research Incubation Fund of Baoshan University(BYPY202216), and the Fundamental Research Funds for Central Universities, China (Grant No. CCNU19QN029), National Natural Science Foundation of China (Grant Nos. 11505071, 61702207, and 61873104), and 111 Project (Grant No. BP0820038.

\nocite{*}

\bibliography{apssamp}

\end{document}